\documentclass[12pt]{article}
\usepackage{graphicx}
\usepackage{epsfig}

\catcode`@=11
\newbox\Lbox
\newbox\Rbox
\newdimen\Twofigdimen
\Twofigdimen=\textwidth
\advance\Twofigdimen by -0.25truein
\divide\Twofigdimen by 2
\def\Lcaption[#1]#2{%
   \stepcounter{figure}%
   \def\@currentlabel{\thefigure}%
   \addcontentsline{lof}{figure}{\numberline{\thefigure}{\ignorespaces #1}}
   \setbox\Lbox=\vtop{\hsize=3in
      \raggedright\tolerance=10000
      \noindent Figure~\thefigure: #2}%
}
\def\Rcaption[#1]#2{%
   \stepcounter{figure}%
   \def\@currentlabel{\thefigure}%
   \addcontentsline{lof}{figure}{\numberline{\thefigure}{\ignorespaces #1}}
   \setbox\Rbox=\vtop{\hsize=3in
      \raggedright\tolerance=10000
      \noindent Figure~\thefigure: #2}%
   \bigskip
   \centerline{\copy\Lbox\hfill\copy\Rbox}%
}
\def\Twofigs#1#2{%
   \hbox to \textwidth{\hbox to \Twofigdimen{\hss{#1}\hss}\hfil
      \hbox to \Twofigdimen{\hss{#2}\hss}}}
\catcode`@=12
\def\({ \left( }
\def\){ \right) }
\def\b{\begin{equation}}
\def\e{\end{equation}}
\def\={\ =\ }

\def\+{\ +\ }
\def\-{\ -\ }

\def\Ls{\cal L \rm}

\def\mumu{$\mu^+\mu^-$}
\def\ee{$e^+e^-$}
\def\pp{$pp$}
\def\ppbar{$p\bar{p}$}

\def\simge{                       
 \mathrel{\rlap{\raise 0.511ex
	\hbox{$>$}}{\lower 0.511ex \hbox{$\sim$}}}}
\def\simle{
    \mathrel{\rlap{\raise 0.511ex
	\hbox{$<$}}{\lower 0.511ex \hbox{$\sim$}}}}

\def\mumu{$\mu ^+\mu ^-$\ }

\def\E0c{\frac{E_0}{c}}
\def\1s2{\frac{1}{\sqrt{2}}}

\begin{document}
\title{HIGH ENERGY COLLIDERS}
\author{R. B. Palmer, J. C. Gallardo\\
\\
Center for Accelerator Physics\\
Brookhaven National Laboratory \\  
Upton, NY 11973-5000, USA}
\maketitle    
\abstract{
We consider the high energy physics advantages, disadvantages 
and luminosity requirements of hadron (\pp, \ppbar), lepton (\ee, \mumu) 
and photon-photon colliders. Technical problems in obtaining  increased energy 
in each type of machine are presented. The machines relative size are also discussed. }

\section{Introduction}

\begin{figure}[!hbt]
\begin{minipage}{0.45\linewidth} 
\centerline{\epsfig{file=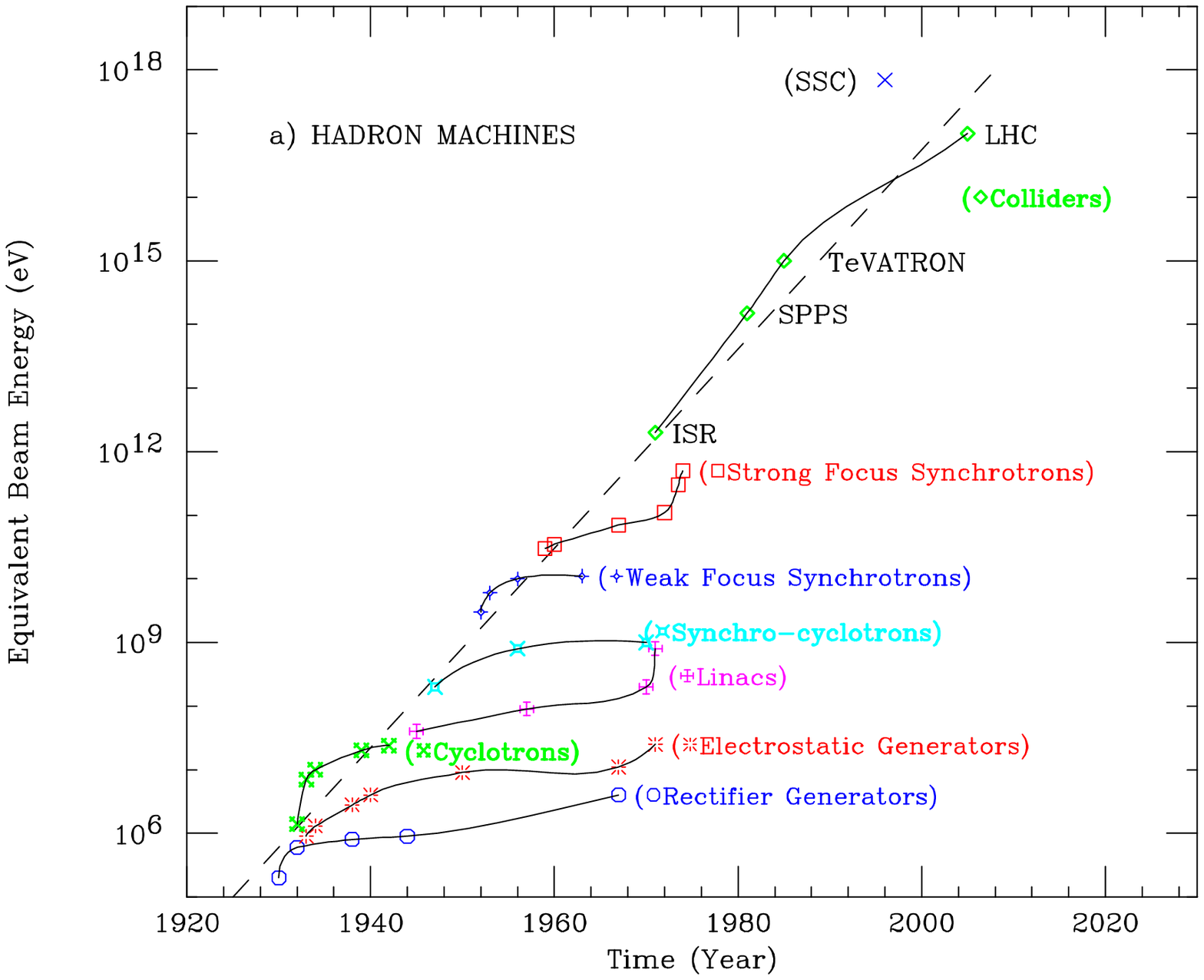,height=4.0in,width=\linewidth}}
\end{minipage}\hfill
\begin{minipage}{0.45\linewidth}
\centerline{\epsfig{file=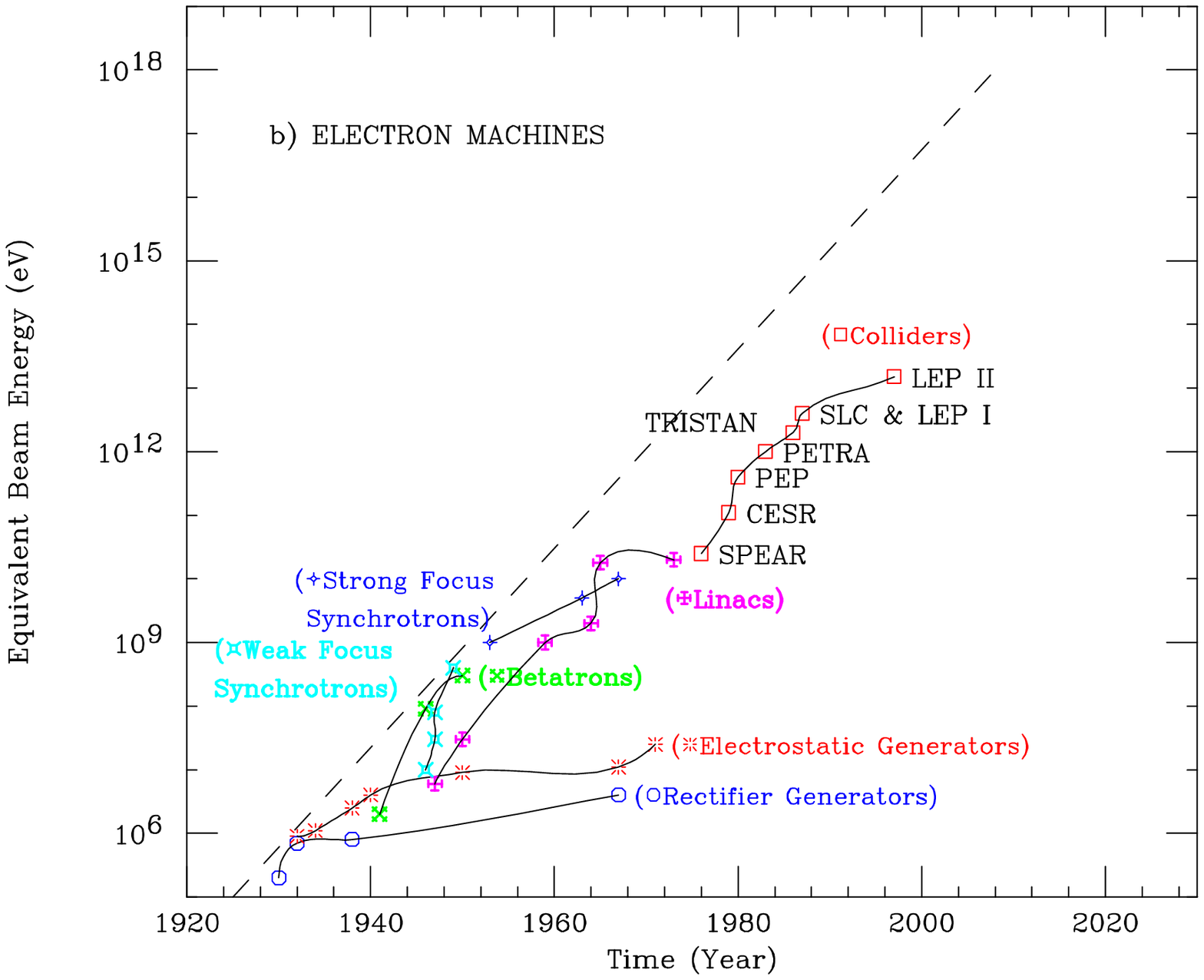,height=4.0in,width=\linewidth }}
\end{minipage}
 \caption{The Livingston Plots: Equivalent beam energy of colliders versus the 
year of their introduction; (a) for Hadron Machines and (b) for Lepton 
Machines.
 \label{liv} }
\end{figure}

   Particle colliders are only the last evolution of a long history of devices used to study the violent collisions of particles on one another. Earlier versions used accelerated beams impinging on fixed targets.
Fig. \ref{liv} shows the equivalent beam energy of such machines, plotted versus the 
year of their introduction. The early data given was taken from the original 
plot by Livingston\cite{livingston}. For hadron, i.e. proton or proton-antiproton, machines (Fig. \ref{liv}a), it 
shows an increase from around $10^{5}$ eV with a rectifier generator in 1930, 
to $10^{15}$ eV at the  Tevatron (at Fermilab near Chicago) in 1988. This 
represents an increase of more than a factor of about 33 per decade (the Livingston line, shown as the dash-line) over 6 decades. By 2005 we expect to have the Large Hadron Collider (at 
CERN, Switzerland) with an equivalent beam energy of $10^{17}$ eV, 
which will almost exactly continue this trend. The SSC, had we built it on 
schedule, would, by this extrapolation, have been a decade too early ! 

The rise in energy of electron machines shown (Fig. \ref{liv}b) is slightly 
less dramatic; but, as we shall discuss below, the relative effective physics 
energy of lepton machines is greater than for hadron machines, and thus
the effective energy gains for the two types of machine 
are comparable.

   These astounding gains in energy ($\times\ 10^{12}$) have been partly bought 
by greater expenditure: increasing from a few thousand dollars for the 
rectifier, to a few billion dollars for the LHC ($\times\ 10^{6}$). The 
other factor ($\times\ 10^{6}$) has come from new ideas. Linear \ee, 
$\gamma - \gamma$, and \mumu colliders are new ideas that we hope will 
continue this trend, but it will not be easy.
\noindent

\section{Physics Considerations}

\subsection{General.}
Hadron-hadron colliders (\pp \, or \ppbar) generate interactions between  the
many constituents of the  hadrons (gluons, quarks and antiquarks); the initial 
states are not defined and most interactions occur at relatively  low energy,
generating a very large background of uninteresting  events. The rate of the
highest energy events is a little higher for antiproton-proton machines, because  
the antiproton contains valence antiquarks that can annihilate on the quarks 
in the proton. But this is a small effect for  colliders above a few TeV, when 
the interactions are dominated by interactions between quarks and antiquarks 
in their seas, and between the gluons. In either case the individual parton-
parton interaction energies (the energies used for physics) are a relatively 
small fraction of the total center of mass energy. This is a disadvantage when 
compared with lepton machines. An advantage, however, is that all final states 
are accessible. In addition, as we saw in Figs.~\ref{liv}, hadron machines have been available with higher energies than lepton devices, and, as a result, most initial  discoveries in Elementary Particle 
Physics have been made with hadron machines. 

In contrast, lepton-antilepton collider generate interactions between the 
fundamental point-like  constituents in their beams, the reactions produced 
are  relatively simple to understand, the full machine energies are available 
for physics and there is negligible background of low  energy events. If 
the center of mass energy is set equal to the  mass of a suitable state of 
interest, then there can be a large  cross section in the {\bf s}-channel, in 
which a single state is  generated by the interaction. In this case, the mass 
and quantum  numbers of the state are constrained by the initial beams. If the  
energy spread of the beams is sufficiently narrow, then precision  
determination of masses and widths are possible. 

A gamma-gamma collider, like the lepton-antilepton machines, would have 
all the machine energy  available for physics, and would have well defined 
initial states, but these states would be different from those with the lepton 
machines, and thus be complementary to them. 

For most purposes (technical considerations aside) \ee and \mumu  colliders
would be equivalent. But in  the particular case of {\bf s}-channel Higgs boson
production,  the cross section, being proportional to the mass squared, is 
more than 40,000 times greater for muons than electrons.  When technical
considerations are included, the situation is  more complicated. Muon beams are
harder to polarize and muon  colliders will have much higher backgrounds from
decay  products of the muons. On the other hand muon collider  interactions
will require less radiative correction and will have  less energy spread from
beamstrahlung.

Each type of collider has its own advantages and disadvantages 
for High Energy Physics: they would be complementary. 
\subsection{Required Luminosity for Lepton Colliders.} 

   In lepton machines the full center of mass of the leptons is 
available for the final state of interest and a ``physics 
energy" $E_{\rm phy}$ can be defined that is equal to the total center of mass 
energy. 
 \b
E_{\rm phy}\ =\ E_{c\ of\ m}
 \e

   Since fundamental cross sections fall as the square of the 
center of mass energies involved, so, for a given rate of events, 
the luminosity of a collider must rise as the square of its 
energy. A reasonable target luminosity is one that would give 
10,000 events per unit of R per year (the cross section for lepton pair 
production is one R, the total cross section is about 20 R, and somewhat 
energy dependent as new channels open up): 
\b
\Ls_{req.}\ \approx \  10^{34}\ (cm^{-2} s^{-1})\  \left( {E_{phy} \over 1\ 
(TeV)} \right)^2 \label{reqlum} 
\e

\subsection{The Effective Physics Energies of Hadron Colliders.}
Hadrons, being composite, have their energy divided between 
their various constituents. A typical collision of 
constituents will thus have significantly less energy than that 
of the initial hadrons. Studies done in Snowmass 82 and 96 
suggest that, for a range of studies, and given the required luminosity (as 
defined in Eq.~\ref{reqlum}), then the hadron machine's effective 
``physics" energy is between about  1/3 and 1/10 of its total. We will take a 
value of 1/7: 
  $$
E_{\rm phy}(\Ls=\Ls_{req.})\ \approx\ {E_{c\ of\ m} \over 7}
  $$
The same studies have also concluded that a factor of 10 in 
luminosity is worth about a factor of 2 in effective physics energy, this 
being approximately equivalent to: 
  $$
E_{\rm phy}(\Ls)\= E_{\rm phy}(\Ls=\Ls_{req.})\ 
\({ \Ls \over \Ls_{req} }\)^{0.3} 
  $$
From which, with Eq.~\ref{reqlum}, one obtains:
 \b
E_{\rm phy} \approx  
  \( {E_{c\ of\ m}\over 7 (TeV)} \)^{0.6}
               \( {\Ls \over 10^{34}(cm^{-2}s^{-1})} \)^{0.2}  (TeV)
\label{Eeffeq}
 \e
\begin{table}[thb!]  
\centering \protect
\caption{Effective Physics Energy of Some Hadron Machines}
\begin{tabular}{lccc}
\hline 
Machine & C of M Energy & Luminosity & Physics Energy \\
         &  TeV         &$cm^{-2}s^{-1}$& TeV   \\
\hline
ISR & .056 & $10^{32}$ & 0.02 \\
Tevatron & 1.8 & $7\times 10^{31}$ & 0.16 \\
LHC   & 14 & $10^{34}$ & 1.5 \\
VLHC   & 60 & $10^{34}$ & 3.6 \\
\hline
\end{tabular}
\label{physE}
\end{table}
Table ~\ref{physE} gives some examples of this approximate ``physics" energy.
It must be emphasized that this effective physics energy is not a well 
defined quantity. It should depend on the physics being studied. The initial 
discovery of a new quark, like the top, can be made with a significantly lower 
``physics" energy than that given here. And the capabilities of different types 
of machines have intrinsic differences. The above analysis is useful only in 
making very broad comparisons between machine types.
\section{Hadron-Hadron Machines}

 \subsection{Luminosity.}
An antiproton-proton collider requires only one ring, compared  with the two
needed for a proton-proton machine (the antiproton has the opposite charge to 
the proton and can thus rotate in the same magnet ring in the opposite 
direction - protons going in opposite directions require two rings with 
bending fields of the opposite sign), but the  luminosity of an antiproton-
proton collider is limited by the  constraints in antiproton production. A 
luminosity of at least $10^{32}\ {\rm cm}^{- 2}{\rm s}^{-1}$ is expected at 
the antiproton-proton Tevatron; and a luminosity of $10^{33}\ {\rm cm}^{- 
2}{\rm s}^{-1}$ may be achievable, but the LHC, a proton-proton machine, is 
planned to have a luminosity  of $10^{34}\ {\rm cm}^{-2}{\rm s}^{-1}$: an order of magnitude higher. Since 
the required luminosity rises with energy, proton-proton machines seem to be 
favored for future hadron colliders. 

   The LHC and other future proton-proton machines might\cite{lumlim}  
 be upgradable to $10^{35}\ {\rm cm}^{-2}{\rm s}^{-1}$, but radiation damage 
to a detector would then be a severe problem. The 60 TeV Really 
Large Hadron Colliders (RLHC: high and low field versions) discussed at 
Snowmass are  being designed as proton-proton machines with luminosities of  
$10^{34}\ {\rm cm}^{-2}{\rm s}^{-1}$ and it seems reasonable to assume that 
this is the highest practical value. 
\subsection{Size and Cost.}
The size of hadron-hadron machines is limited by the field of  the magnets used
in their arcs. A cost minimum is obtained when a  balance is achieved between
costs that are linear in length, and  those that rise with magnetic field. The
optimum field will  depend on the technologies used both for the the linear 
components (tunnel, access, distribution, survey, position  monitors,
mountings, magnet ends, etc) and those of the magnets  themselves, including
the type of superconductor used. 

The first hadron collider, the 60 GeV ISR at CERN, used  conventional iron pole
magnets at a field less than 2~T. The  only current hadron collider, the 2 TeV
Tevatron, at FNAL, uses NbTi  superconducting magnets at approximately
$4\,{}^\circ K$ giving a bending field of about 4.5 T. The  14 TeV Large 
Hadron Collider (LHC), under construction at CERN,  plans to use the same 
material at $1.8\,{}^\circ K$ yielding bending fields of about $8.5\,{\rm T}.$ 

Future colliders may use new materials allowing even higher  magnetic fields.
 Model magnets have been made with ${\rm Nb_3Sn},$ and  studies are underway on 
the use of high T$_c$ superconductor.  {Bi$_2$Sr$_2$Ca$_1$Cu$_2$O$_8$} (BSCCO) 
material is currently available in useful  lengths as powder-in-Ag tube 
processed tape. It has a higher  critical temperature and field than 
conventional superconductors,  but, even at $4\,{}^\circ K,$ its current 
density is less than ${\rm Nb_3Sn}$  at all fields below 15~T. It is thus 
unsuitable for most accelerator magnets. In contrast YBa$_2$Cu$_3$O$_7$ (YBCO) 
material has a current density above that for Nb$_3$Sn ($4\,{}^\circ K$), at 
all fields and temperatures below $20\,{}^\circ K.$  But this material must be 
deposited on specially treated  metallic substrates and is not yet available 
in lengths greater  than 1~m. It is reasonable to assume, however, that it 
will be available in useful lengths in the not too distant future.  
it mean for hadron colliders. 

\begin{figure}[htb!] 
\centerline{\epsfig{file=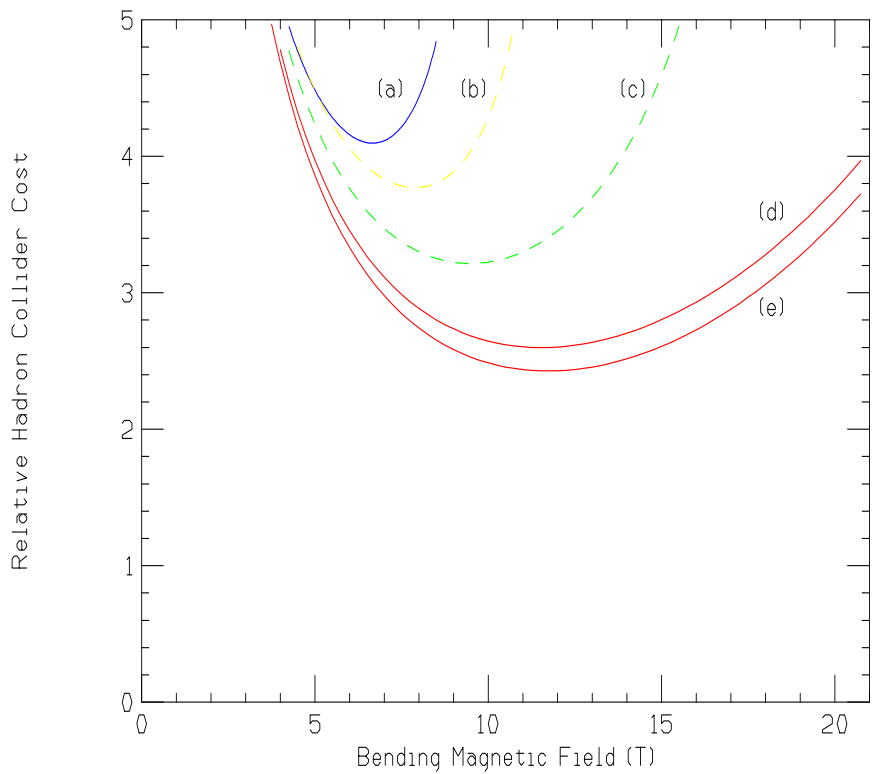,height=4.5in,width=4.0in,angle=90}}
\caption{Relative 
costs of a collider as a function of its bending magnetic field, for different 
superconductors and operating temperatures. \label{mag} } 
\end{figure} 
A parametric study was undertaken to learn what the use of  such materials might 
do for the cost of colliders. 2-in-1 cosine  theta superconducting magnet 
cross sections (in which the two magnet coils are circular in cross section, 
have a cosine theta current distributions and are both enclosed in a single 
iron yoke) were calculated using  fixed criteria for margin, packing fraction, 
quench protection,  support and field return. Material costs were taken to be 
linear  in the weights of superconductor, copper stabilizer, aluminum  
collars, iron yoke and stainless steel support tube. The  cryogenic costs were 
taken to be inversely proportional to the operating temperature, and linear in 
the outer surface area of  the cold mass. Tunnel, access, vacuum, alignment, focusing, and diagnostic costs were taken to be linear with tunnel length. The relative values of the cost dependencies 
were scaled from LHC  estimates. 

Results are shown in Fig.~\ref{mag}. Costs were calculated  assuming NbTi at
(a) $4\,{}^\circ K$, and (b) $1.8\,{}^\circ K,$  Nb$_3$\,Sn at (c) 
$4\,{}^\circ K$ and YBCO High T$_c$ at $20\,{}^\circ K$ (d) and (e).  
NbTi  and Nb$_3$\,Sn  costs per unit weight were taken to be the  same; YBCO
was taken to be either equal to NbTi (in (d)), or 4 times NbTi (in (e)).
It is seen that the optimum field moves from about 6~T for  NbTi at
$4\,{}^\circ K$ to about 12~T for YBCO at $20\,{}^\circ K$;  while the total
cost falls by almost a factor of 2.

One may note that the optimized cost per  unit length remains approximately 
constant. This might have been  expected: at the cost minimum, the cost of 
linear and field  dependent terms are matched, and the total remains about 
twice  that of the linear terms.  

The above study assumes this particular  type of magnet and
may not be indicative of the optimization for  radically different designs. A
group at FNAL\cite{pipe} is  considering an iron dominated, alternating
gradient, continuous,  single turn collider magnet design (Low field RLHC). Its
field would  be only 2~T and circumference very large (350 km for 60 TeV), but 
with its simplicity and with tunneling innovations, it is hoped to  make its
cost lower than the smaller high field designs. There  are however greater
problems in achieving high luminosity with  such a machine than with the higher
field designs. 
 \section{Circular \ee Machines}

\subsection{Luminosity.}
 The luminosities of most circular electron-positron  colliders have been 
between $10^{31}$ and $10^{32}\,{\rm cm}^{-2}{\rm s}^{-1}$,  CESR is fast approaching 
$10^{33}\,{\rm cm}^{-2}{\rm s}^{-1}$ and machines are  now being constructed 
with even higher values. Thus, at least in  principle, luminosity does not seem 
to be a limitation (although it may be noted that the 0.2~TeV electron-
positron collider LEP has a  luminosity below the requirement of 
Eq.\ref{reqlum}). 

  \subsection{Size and Cost.}
At energies below 100 MeV, using a reasonable bending  field, the size and cost 
of a circular electron machine is  approximately proportional to its energy. 
But at higher energies,  if the bending field $B$ is maintained, the energy 
lost $\Delta  V_{\rm turn}$ to synchrotron radiation rises rapidly 
 \b
 \Delta  V_{\rm turn}\ 
\propto \ {E^4 \over R\ m^4}\ \propto \ {E^3\ B \over m^4}
\label{syncheq}
 \e
and soon becomes excessive ($R$ is the radius of the ring). A cost minimum is 
then obtained when the cost of the ring is balanced by the cost of the  rf 
needed to replace the synchrotron energy loss. If the ring  cost is 
proportional to its circumference, and the rf is  proportional to its voltage 
then the size and cost of an  optimized machine rises as the square of its 
energy.

   The highest energy circular \ee collider is the LEP at CERN which has a 
circumference of 27 km, and will achieve a maximum center of mass 
energy of about 0.2 TeV. Using the predicted scaling, a 0.5 TeV 
circular collider would have to have a 170 km circumference, and 
would be very expensive. 
 \section{\ee Linear Colliders}

For energies much above that of LEP (0.2 TeV) it is probably
impractical to build 
a circular electron collider. The only  possibility then
is to build two electron 
linacs facing one  another. Interactions occur at the center, and the 
electrons,  after they have interacted, must be discarded. 
The size of such colliders is now dominated by the length of the two linacs and is inversely proportional to the average accelerating gradient in those structures. In current proposals\cite{bluebook} using conventional rf, these lengths are far greater than the circumferences of hadron machines of the same beam energy, but as noted in section 2.3, the effective physics energy of a lepton machine is higher than that of a hadron machine with the same beam energy, thus offsetting some of this disadvantage.
\subsection{Luminosity.}
 The luminosity $\Ls$ of a linear collider can be written: 
 \b
   \label{lumeq}
   \Ls \= {1 \over 4\pi E}\ \ {N \over \sigma_x}\ 
    {P_{\rm beam} \over \sigma_y}
   \ \ n_{\rm collisions}
 \e
where $\sigma_x$ and 
$\sigma_y$ are average beam spot sizes including any pinch 
effects, and we take $\sigma_x$ to be much 
greater than $\sigma_y$. $E$ is the 
beam energy, $P_{\rm beam}$ is the total beam power,
and, in this case, $n_{\rm collisions}=1$.
This can be expressed\cite{yokoyachen} as,
 \b
\label{constraint}
\Ls \ \approx 
       \ {1 \over 4\pi E}\ \ {n_\gamma \over 2 r_o \alpha \ U(\Upsilon)}\ \
       \ \ {P_{\rm beam} \over \sigma_y}
 \e
where the quantum correction $U(\Upsilon)$ is given by 
 \b
   U(\Upsilon)\ \approx \ \sqrt{{1 \over {1+\Upsilon^{2/3}}}  }
 \e
with
 \b
   \Upsilon\ \approx \ {2 F_2 r_o^2 \over \alpha} 
   \ {N \ \gamma \over \sigma_z \ \sigma_x}
 \e
$F_2\approx 0.43$, $r_o$ is the classical 
electromagnetic radius, $\alpha$ is 
the fine-structure constant, and $\sigma_z$ is the  {\it rms} bunch length. 
The quantum correction $\Upsilon$
is close to unity for all 
proposed machines with energy less than 2 TeV, and this term is often 
omitted\cite{peskin}. 
Even in a 5 TeV design\cite{me}, an $\Upsilon$ of 21
gives a suppression factor of only 3. 
 $n_{\gamma}$ is the 
number of photons emitted by one electron as it passes through the 
other bunch. 
If $n_\gamma$ is significantly greater than one, then problems are encountered 
with 
backgrounds of electron pairs and mini-jets, or
with unacceptable beamstrahlung energy loss.
Thus $n_\gamma$ 
can be taken as a rough criterion of these effects and
constrained to a fixed value. We then find:
 $$
\Ls  \ \propto {1 \over  E}\ \  {P_{beam} \over \sigma_y\ U(\Upsilon)}
 $$
which may be compared to the required luminosity that increases 
as the square of energy, giving the requirement: 
 \b
\label{Ecubed}
{P_{\rm beam} \over \sigma_y\ U(\Upsilon)}\  \propto \ E^3.
 \e
It is this requirement that makes it hard to design very high 
energy linear colliders. High beam power demands high efficiencies and heavy 
wall power consumption. A small $\sigma_y$ requires tight tolerances, low beam 
emittances and strong final focus. And a small value of $U(\Upsilon)$ is hard 
to obtain because of its weak dependence on $\Upsilon$ ($\propto \Upsilon^{-
1/3}$). 

\subsection{Conventional RF.}

 The gradients for structures have limits that are frequency dependent, but the real limit on accelerating gradients in these designs come from a trade 
off between the cost of rf power against the cost of length. The use of high 
frequencies reduces the stored energy in the cavities, reducing the rf costs 
and allowing higher accelerating gradients: the optimized gradients being 
roughly proportional to the frequency up to a limit of approximately $250\,{\rm MeV/m}$ at a frequency of the order of $100\,{\rm GHz}.$  One might thus conclude then that 
higher frequencies should always be preferred. There are however counterbalancing 
considerations from the requirements of luminosity. 
 \begin{figure}[hbt!] 
\centerline{\epsfig{file=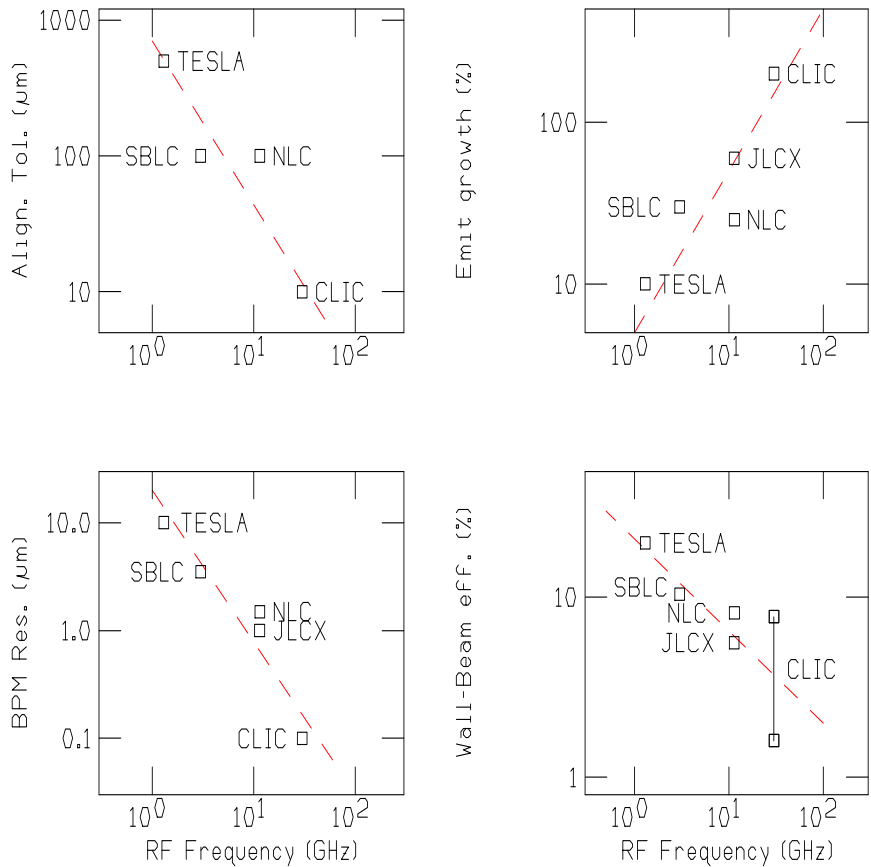,height=4.5in,width=4.5in}}
 \caption{Dependence of some sensitive parameters of 0.5 TeV proposed linear colliders
as a function of their rf frequencies.
 \label{freq} }
 \end{figure} 
    
Fig.~\ref{freq}, using parameters from current 0.5 TeV linear collider 
proposals \cite{bluebook}, plots some relevant parameters against 
the rf frequency. One sees that as the frequencies rise, 
 \begin{itemize}
 \item the required alignment tolerances get tighter;
 \item the resolution of beam position monitors must also be better; and
 \item despite these better alignments, the calculated  emittance growth 
during acceleration is worse; and
 \item the wall-power to beam-power efficiencies are also less.
 \end{itemize}
Thus while length and cost considerations may favor high 
frequencies, yet luminosity considerations would prefer lower 
frequencies. 
\subsection{Superconducting RF.}

         If, however, the rf costs can be reduced, for instance when 
superconducting  cavities are used, then there will be no trade off between rf 
power cost and length and higher gradients would lower the 
length and thus the cost.  Unfortunately the gradients achievable 
in currently operating niobium superconducting cavities is lower than that 
planned in the higher frequency conventional rf colliders. Theoretically the 
limit is about 40~MV/m, but practically 25~MV/m is as high as seems possible. 
Nb$_3$Sn and high Tc materials may allow higher field gradients in the future. 

The removal of the requirements for very high peak rf power allows 
the choice of longer wavelengths (the TESLA collaboration is proposing 23 cm 
at 1.3 GHz) thus greatly relieving the emittance requirements and tolerances, for a given luminosity. 

At the current 25 MeV per meter gradients, the length and cost of a 
superconducting machine is probably higher than for the conventional rf designs. With
greater luminosity more certain, its proponents can argue that it is worth the greater price.
If, using new superconductors, higher gradients become possible, thus reducing lengths and costs,
then the advantages of a superconducting 
solution might become overwhelming.   
\subsection{At Higher Energies.}
\begin{figure}[ht!] 
\centerline{\epsfig{file=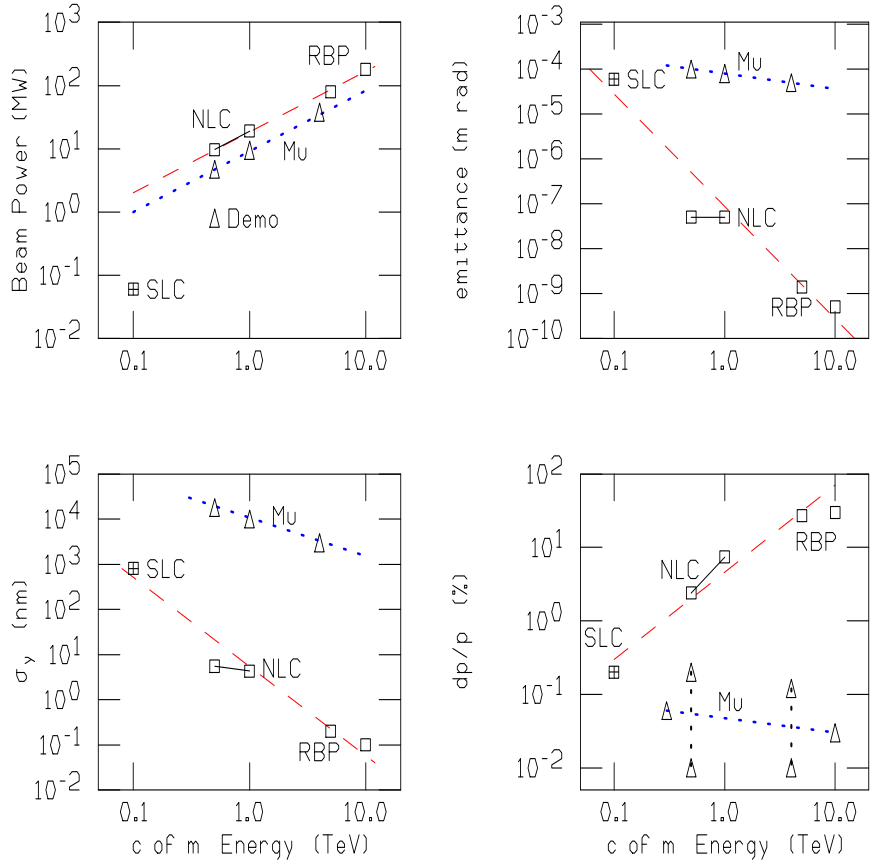,height=4.5in,width=4.5in}}
 \caption{Dependence of some sensitive 
parameters on linear collider energy, with comparison of same parameters for 
\mumu colliders. 
 \label{energy} }
 \end{figure}

    At higher energies (as expected from Eq.~\ref{Ecubed}),  
obtaining the required luminosity gets harder. Fig.\ref{energy} 
shows the dependency of some example machine 
parameters with energy.  SLC is taken as the example at 0.1 TeV, 
NLC parameters at 0.5 and 1 TeV, and 5 and 10 TeV examples are 
taken from a review paper by one of the authors\cite{me}. One sees 
that: 
 \begin{itemize}
 \item the assumed beam power rises approximately as $E$;
 \item the vertical spot sizes fall approximately as $E^{-2}$;
 \item the vertical normalized emittances fall even faster: $E^{-2.5}$; 
and
 \item the momentum spread due to beamstrahlung has been allowed to rise. 
 \end{itemize}

   These trends are independent of the acceleration method, 
frequency, etc, and indicate that as the energy and required 
luminosity rise, so the required beam powers, efficiencies, 
emittances and tolerances will all get harder to achieve. The 
use of higher frequencies or exotic technologies that would allow 
the gradient to rise, will, in general, make the achievement of 
the required luminosity even more difficult. It may well prove
impractical to construct linear electron-positron colliders, with 
adequate luminosity, at energies above a few TeV. 
 \section{$\gamma-\gamma$ Colliders}

   A gamma-gamma collider\cite{telnov} would use opposing electron 
linacs, as in a linear electron collider,  but just prior to the 
collision point, laser beams would be Compton backscattered off the 
electrons to generate photon beams that would collide at the IP instead of the 
electrons. If suitable geometries are used, the mean 
photon-photon energy could be 80\% or more of that of the 
electrons, with a luminosity about 1/10th. 

   If the electron beams, after they have backscattered the 
photons, are deflected, then backgrounds from beamstrahlung can be eliminated. 
The constraint on ${N / \sigma_x}$ in Eq.\ref{lumeq} is thus removed and one 
might hope that higher luminosities would now be possible by raising $N$ and 
lowering $\sigma_x$. Unfortunately, to do this, one needs sources of 
bunches with larger 
numbers of electrons and smaller emittances, and one must find ways to 
accelerate and focus such  beams without excessive emittance growth. 
Conventional damping rings will have difficulty doing this\cite{mygamma}. 
Exotic electron sources would be needed, and methods using lasers to 
generate\cite{palmerchen} or cool\cite{telnovcool} 
 the electrons and positrons are under 
consideration.  

 Clearly, although gamma-gamma collisions can and should be made 
available at any future electron-positron linear collider, to add 
physics capability, whether they can give higher luminosity for a 
given beam power is less clear.
\section{\mumu Colliders}

\subsection{Advantages and Disadvantages}
The possibility of muon colliders was introduced by Skrinsky et 
 al.\cite{ref2} and Neuffer\cite{ref3} and has been aggressively 
 developed over the past two years in a series of meetings and 
 workshops\cite{ref4,ref5,ref6,ref7,book}. 

   The main advantages of muons, as opposed to 
electrons, for a lepton collider are:
 \begin{itemize}
\item
   The synchrotron radiation, that forces high energy electron 
colliders to be linear, is (see Eq. \ref{syncheq}) inversely 
proportional to the fourth power of mass: It is negligible in muon 
colliders. Thus a muon collider
can be circular. In practice this means in 
can be smaller. The linacs for the SLAC proposal for a 0.5 TeV Next Linear Collider would be 20 km 
long. The ring for a muon collider of the same energy would be 
only about 1.3 km circumference. 
\item
   The luminosity of a muon collider is given by the same formula 
(Eq.~\ref{lumeq}) as given above for an electron positron collider, but there 
are two significant changes: 1) The classical radius $r_o$ is now that for the 
muon and is 200 times smaller; and 2) the number of collisions a bunch can 
make $n_{collisions}$ is no longer 1, but is now 
limited only by the muon lifetime and becomes related to the average 
bending field in the muon collider ring, with 
 $$
n_{collisions} \ \approx \ 150  \ B_{ave}
 $$
With an average field of 6 Tesla, $n_{collisions}\approx 900$.
These two effects give muons an {\it in principle} luminosity advantage of 
more than $10^5$. As a result, for the same luminosity, the required beam 
power, spot sizes, emittances and energy spread are far less in \mumu 
colliders than in \ee machines of the same energy. The comparison is made in 
Fig.~\ref{energy} above. 

\item The suppression of synchrotron radiation induced by the opposite bunch 
(beamstrahlung) allows the use of beams with lower momentum 
spread, and QED radiation is reduced.
\item $s$-channel Higgs production is enhanced by a factor of $(m_\mu/m_e)^2
\approx 40000$. This combined with the lower momentum spreads would allow 
more precise determination of Higgs masses, widths and branching ratios. 

 \end{itemize} 

But there are problems with the use of muons:
 \begin{itemize}
 \item Muons can be obtained from the decay of pions, made by higher energy 
protons impinging on a target. But in order to obtain enough muons, a high 
intensity proton source is required with very efficient capture of the pions, 
and muons from their decay. 
 \item The selection of fully polarized muons is inconsistent with the 
requirements for efficient collection. Polarizations only up to 50 \% are 
practical, and some loss of luminosity is inevitable (\ee machines can 
polarize the $e^-$'s up to $\approx$ 85 \%). 
 \item Because the muons are made with very large emittance, 
they must be cooled, and this must be done very rapidly because of their short 
lifetime. Conventional synchrotron, electron, or stochastic cooling is too 
slow. Ionization cooling\cite{ioncool} is the only clear possibility, but does not cool to 
very low emittances. 
 \item Because of their short lifetime, conventional 
synchrotron acceleration would be too slow. Recirculating 
accelerators or pulsed synchrotrons must be used.
 \item
Because they decay while stored in the collider, muons radiate
the ring and 
detector with decay electrons. Shielding is essential and backgrounds
will be high.
 \end{itemize}
\subsection{Design Studies} 
\begin{figure}
\centerline{\epsfig{file=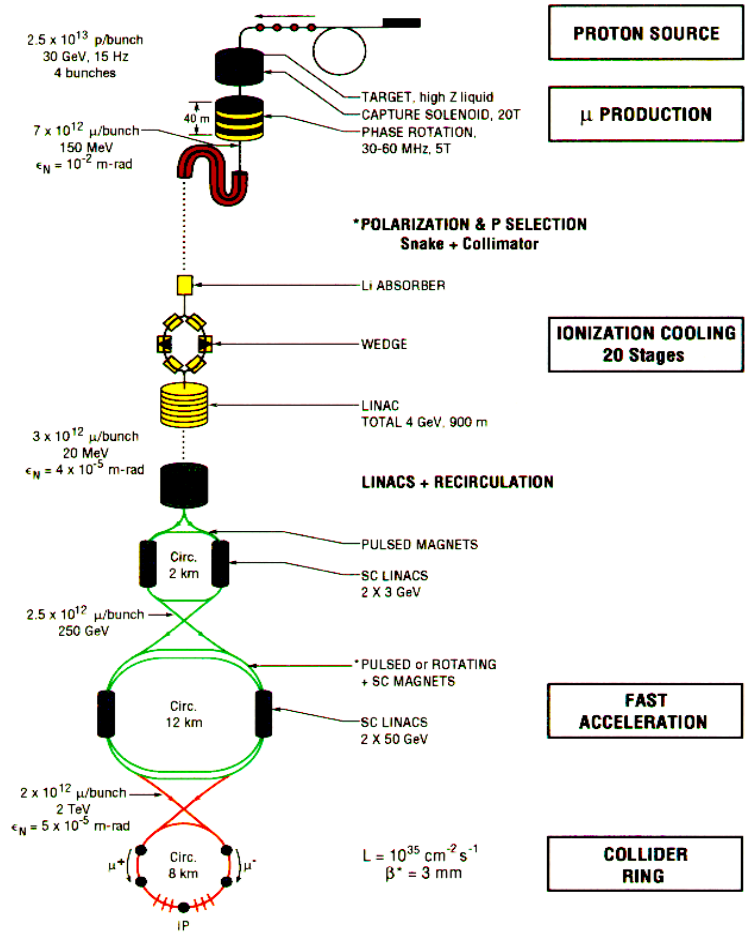,height=5.0in,width=4.0in}}
 \caption{Overview of a 4 TeV Muon Collider
 \label{overview}}
 \end{figure}

\begin{figure}[hbt!]
\centerline{\epsfig{file=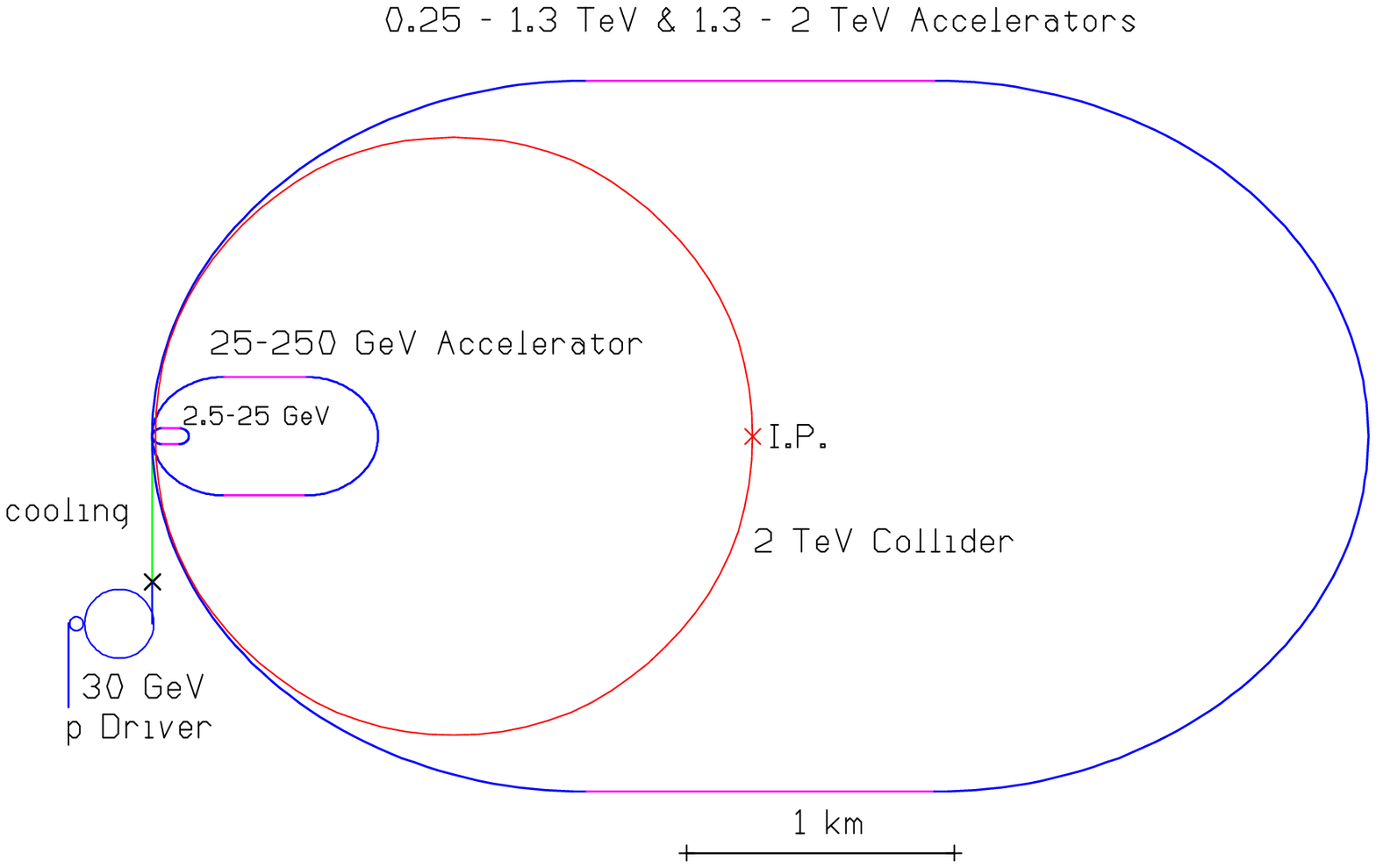,height=3.5in,width=5.5in}}
\caption{Layout of the collider and accelerator rings.
 \label{accpict}}
 \end{figure}

A collaboration, lead by BNL, FNAL and LBNL,  with contributions from 
18 institutions has been studying a 4 TeV, high luminosity  scenario 
and presented a Feasibility Study\cite{book} to the  1996 Snowmass 
Workshop. The basic parameters of this machine are shown 
schematically in Fig.~\ref{overview} and given in Table ~\ref{sum}.
Fig.~\ref{accpict} shows a possible layout of such a machine. 

Table ~\ref{sum} also gives the parameters of
a 0.5 TeV demonstration machine based on the AGS as an 
injector. It is assumed that a demonstration version based on upgrades 
of the FERMILAB, or CERN machines would also be possible. 
\begin{table}[thb!]  
\centering \protect
\caption{Parameters of Collider Rings
\label{sum}}
\begin{tabular}{llcc}
\hline
c-of-m Energy              & TeV      & 4     & .5       
\\
\hline
Beam energy                & TeV      &     2    &   .25 
   \\
Beam $\gamma$              &          &   19,000 &  
2,400   \\
Repetition rate            & Hz       &    15    &    
2.5    \\
Proton driver energy       & GeV      & 30 & 24 \\
Protons per pulse          &          &  $10^{14}$ & 
$10^{14}$\\
Muons per bunch            &   &   2 $10^{12}$    &    4 $10^{12}$  
    \\
Bunches of each sign       &          &   2      &    1  
    \\
Beam power                 & MW       &   38     &   .7 
\\
Norm. {\it rms} emit.   $\epsilon_n$   &$\pi$ mm mrad  & 
 50  &  90    \\
Bending Fields              &  T   &    9    &    9     
\\
Circumference              &  Km      &    8    &    1.3 
   \\
Ave. Bending Fields    & T   & 6    &   5       \\
Effective turns          &       & 900    &   800   \\
$\beta^*$ at intersection   & mm     &   3   &   8     \\
{\it rms} bunch length   & mm     &   3   &   8     \\
{\it rms} I.P. beam size        & $\mu m$&   2.8 &  17   
 \\
Chromaticity             &         & 2000-4000 & 40-80 
\\
$\beta_{\rm max}$ & km   & 200-400 & 10-20 \\
Luminosity &${\rm cm}^{-2}{\rm s}^{-1}$& 
$10^{35}$&$10^{33}$\\
\hline
\end{tabular}
\end{table}

The main components of the 4 TeV collider would be:
\begin{itemize}
\item  A proton source with KAON\cite{kaon} like parameters (30 GeV, $10^{14}$ 
protons per pulse, at 15 Hz). 

\item A liquid metal target surrounded by a 20~T hybrid solenoid to make and capture pions. 

\item A 5 T solenoidal channel to allow the pions to decay into 
muons, with rf cavities to,
at the same 
time, decelerate the fast ones that come first, while 
accelerating the slower ones that come later. Muons from 
pions in the 100-500 MeV range emerge in a 6 m long bunch at 150 
$\pm$ 30 MeV.

\item A solenoidal snake and collimator to select 
the momentum, and thus the polarization, of the muons. 

\item A sequence of 20 ionization cooling stages, each consisting of: 
 a) energy loss material in a strong focusing environment for 
transverse cooling; b) linac reacceleration and c) lithium wedges in 
dispersive environments for cooling in momentum space. 

\item A linac and/or recirculating linac pre-accelerator, followed by
a sequence of pulsed field synchrotron accelerators using 
superconducting linacs for rf.

\item An isochronous collider ring with locally corrected low beta 
($\beta$=3 mm) insertion.
\end{itemize}

\subsection{Status and Required R and D}

Muon Colliders are promising, but they are in a far less developed state than 
hadron or \ee machines. No muon collider has ever been built. Much theoretical and experimental work will be needed 
before one will even know if they are possible. In particular, theoretical 
work is needed on the cooling sequence, on the collider ring, and on 
estimations of background in the detectors. The highest priority experimental 
work is: 
 \begin{itemize}
 \item
Demonstration of ionization cooling;
 \item 
Demonstration of liquid targets, solenoid pion capture and the use of rf near 
such a source; 
 \item
Construction of model pulsed magnets for the accelerator and large aperture 
superconducting quadrupoles for the intersection region of the collider. 
 \end{itemize}
\section{Comparison of Machines}
In Fig.~\ref{length}, the effective physics energies (as defined by 
Eq.~\ref{Eeffeq}) of representative machines 
are plotted against their total tunnel lengths. We note:
 \begin{figure}[hbt!] 
\centerline{\epsfig{file=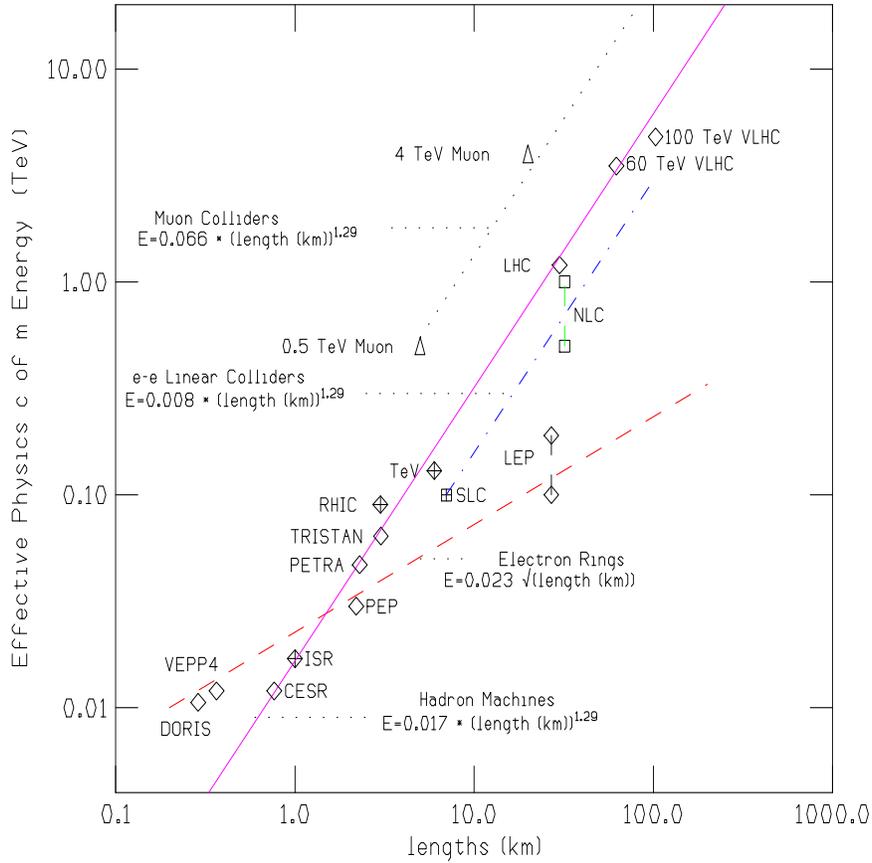,height=4.5in,width=4.5in}}
 \caption{Effective physics energies of 
colliders as a function of their total length.
 \label{length} }
\end{figure} 
 \begin{itemize}
 \item  Hadrons Colliders:
 It is seen that the energies of machines rise with their size, and
that this rise is faster than linear ($E_{\rm eff}\propto L^{1.3}$). This
extra rise is a reflection of the increases in bending magnetic 
field used, as new technologies and materials have become available. 

 \item Circular Electron-Positron Colliders:
The energies of these machines rise approximately as the square root of their 
size, as expected from the cost optimization discussed in section 4 above.

 \item Linear Electron-Positron Colliders:
The SLAC Linear Collider is the only existing machine of this type. One example of a 
proposed machine (the NLC) is plotted. The line drawn has the same slope as 
for the hadron machines and implies a similar rise in accelerating 
gradient, as technologies advance. 

 \item Muon-Muon Colliders:
Only the 4 TeV collider, discussed above, and the 0.5 TeV {\it demonstration
machine}  have been plotted. The line  drawn has the same slope as for the
hadron machines. 
\end{itemize}
            
\begin{figure*}[hbt!] 
\centerline{\epsfig{file=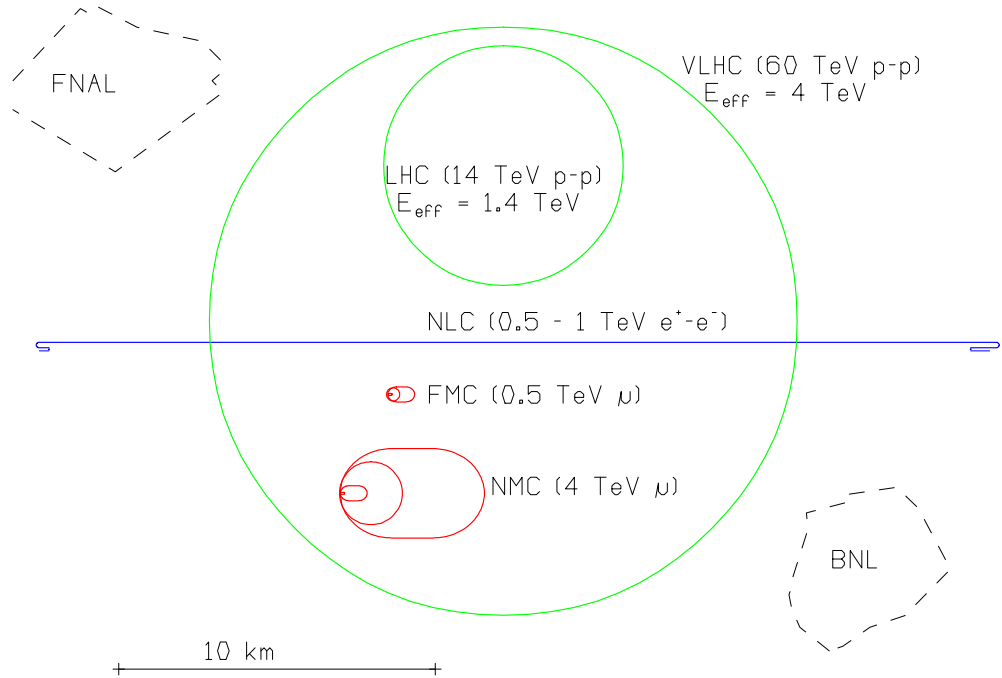,height=5.18in,width=3.52in,angle=90}}
 \caption{Approximate sizes of some possible future colliders.
 \label{examples} }
\end{figure*}

   It is noted that the muon collider offers the greatest energy per unit 
length. This is also apparent in Fig.~\ref{examples}, in which the 
footprints of a number of proposed machines are given on the same scale. 

\section{Conclusions}
Our conclusions, with the caveat that they are indeed only 
our opinions, are: 
 \begin{itemize}
 \item   The LHC is a well optimized and appropriate next step towards high 
{\it effective physics} energy.

\item   A Very Large Hadron Collider with energy greater than the 
SSC (e.g. 60 TeV c-of-m) and cost somewhat less than the SSC, may 
well be possible with the use of high T$_c$ superconductors that 
may become available. 

\item   A ``Next Linear Collider" is the only clean way to 
complement the LHC with a lepton machine, and the only way to do 
so soon. But it appears that even a 0.5 TeV collider may be more 
expensive than the LHC, has significantly less {\it effective physics energy}, and will be technically 
challenging. Obtaining the design luminosity may not be easy. 

\item Extrapolating conventional rf \ee linear colliders to 
energies above 1 or 2 TeV will be very difficult. Raising the rf 
frequency can reduce length and probably cost for a given energy, 
but obtaining  luminosity increasing as the square of energy, as 
required, may not be feasible. 

\item   Laser driven accelerators are becoming more realistic and 
can be expected to have a significantly lower cost per TeV. But 
the ratio of luminosity to wall power is likely to be significantly worse than for 
conventional rf driven machines. Colliders using such technologies are thus 
 unlikely to achieve the very high luminosities needed for physics research at higher energies.

\item A higher gradient superconducting Linac collider using Nb$_3$Sn or
high T$_c$ materials, if it became 
technically possible, could be the most economical
way to attain the required luminosities in a higher energy \ee collider.

\item   Gamma-gamma collisions can and should be obtained at any 
future electron-positron linear collider. They would add physics 
capability to such a machine, but, despite their freedom from the 
beamstrahlung constraint, may not achieve higher 
luminosity.

\item   A Muon Collider, being circular, could be far smaller 
than a conventional rf \ee linear collider of the same 
energy. Very preliminary estimates suggest that it would also be 
significantly less expensive. The ratio of luminosity to wall power for 
such machines, above 2 TeV, may be better than that for 
electron positron machines, and extrapolation to a center of mass 
energy of 4 TeV does not seem unreasonable. If research 
and development can show that it is practical, then a 
0.5 TeV muon collider could be a useful complement to \ee colliders, and, 
at higher energies (e.g. 4 TeV), could be a viable alternative. 
   \end{itemize}

\section{Acknowledgments}
We acknowledge important contributions from many colleagues, especially 
those that contributed to the feasibitity study submitted to the Snowmass 
Workshop 96  Proceedings\cite{book} from which much of the material and some 
text for this report, has been taken: 
In particular we acknowledge the contributions of  the Editors of each one of the chapters of the $\mu^+\mu^-$ Collider: A Feasibility Study: V. Barger, 
J. Norem, R. Noble, H. Kirk, R. Fernow, D. Neuffer, J. Wurtele, D. Lissauer, 
M. Murtagh, S. Geer, N. Mokhov and D. Cline.
\medskip
This research was supported by the U.S. Department of Energy under Contract No.
DE-ACO2-76-CH00016 and DE-AC03-76SF00515.

\section{Discusion }

\noindent R.Taylor:  
I was afraid if it was going to run over but it 
worked well because we spent hardly any time on what is wrong with 
muon colliders compared to the length of time we spent on what was 
wrong with linear colliders. (laugh) 

\noindent K.Henry:  
So what's wrong with the muon colliders? (laugh)

\noindent Palmer:  
I can tell you which parts of the muon 
collider keep me awake at night. That changes, of course,  
from week to week. Enormous progress has been made even in last few 
months. The collider lattice had been a problem, but 
doesn't worry me any more. 
We also had a serious difficulty in the transverse cooling lattices. 
When we first tried tracking particles through, some muons 
never came out. They were hitting resonances. Now we understand that
problem and have tracked through transverse cooling sections that 
work. 

  But we have not done energy cooling yet. We know theoretically how to 
do it, but we haven't got a realistic lattice and tracked muons 
through it. Having been burnt once, I will have sleepless 
nights until we get past that hurdle. 

  The collider ring may have instability problems 
that are not fully understood. We think that it will have to have 
BNS damping applied by using RF quadrupoles, but we haven't worked 
that out. 

  We haven't done many things that need to be done, but I do not yet 
see any insuperable problems. I do not sleep that badly. (laugh)

\noindent Erich Vogt:  
Have you considered using surface muons which have been considered at 
KEK as an alternative muon source? 

\noindent Palmer:  
Yes, but we need bunches with very large 
numbers of muons in order to get luminosity. It seems to be difficult 
to get them from surface muons. And there is a more basic problem, we 
need both charges, I do not think this is possible with surface muons. 

\noindent Edward Witten:  
What fraction of muons decay before entering the ring?

\noindent Palmer:  
With the parameters we've considered, about three-fourths are lost.
Half decay during the cooling sequence, and half of the remainder 
decay during acceleration. 

\noindent Alfred Mann:  
It would be interesting to hear about the shielding problems that 
arise in the muon-muon collider. 

\noindent Palmer:  
I think I know what you're trying to get at. (laugh) 
The radiation from decay electrons in the ring itself 
can be shielded relatively easily. Dumping 2 Tev muons 
is more difficult because it takes
2 km of concrete to stop them, but that too is ok.
The problem you may be hinting at, which I didn't mention because 
we are not yet sure about the calculation, is radiation from 
neutrinos. Muons decaying in a straight section of the collider ring
produce a neutrino beam with opening angle of 
$1/\gamma$ that, for a 4 TeV collider, is only a meter or two wide 35 km away. 
The neutrino cross section is small, but rising linearly with energy, and 
there are 20 mega-watts of power in that beam. The resulting radiation level 
appears to be close to the legal limit.
You can't shield it and it always breaks ground somewhere because 
unfortunately the earth is round. (laugh) 
It rises as the fourth power of the energy  
and is only inverse with the machine depth. Thus, even if a 4 TeV \mumu 
collider is just ok, a 10 Tev collider is probably  impractical.

\noindent Leon Lederman:  
Going back to the beginning of your talk on the Livingston Plot, 
you said that there is a $10^6$ rise of cost rise for a $10^{12} $ rise in 
energy so we are $10^6$ cleverer.(laugh)  
Did you include inflation in those numbers? 

\noindent Palmer:  
Yes, I did, but I may not have done it right. 
Down the bottom I had 100 Kev and 
I said to myself what I could buy that now for 
a few thousand dollars. This is not fair because in 1930 you could not 
buy one and it must have taken quite a bit of labor to build one. I did not 
try and estimate that cost.

\end{document}